\documentclass{article}

\def\diff{\partial}
\def\g{\gamma}
\newcommand{\tg}{\tilde g}
\def\dz{\dot{z}}
\def\dbz{\dot{\bar z}}
\newcommand{\nn}{\nonumber\\}
\newcommand{\p}[1]{(\ref{#1})}
\newcommand{\cZ}{{\cal Z}}

\newcommand{\bB}{{\overline B}}
\newcommand{\bA}{{\overline A}}
\newcommand{\bz}{{\bar z}}
\newcommand{\bp}{{\bar p}}
\newcommand{\hp}{{\hat p}}
\newcommand{\bn}{{\overline\nabla}}
\newcommand{\bhp}{{\hat{ \overline p}}}

\newcommand{\cbZ}{\overline{\cal Z}}

\newcommand{\cF}{{\cal F}}

\newcommand{\cbF}{\overline{\cal F}}
\newcommand{\bF}{{\overline F}}

\newcommand{\bxi}{{\bar\xi}}
\newcommand{\bpsi}{{\bar\psi}}

\newcommand{\ba}{\begin{array}}
\newcommand{\ea}{\end{array}}
\newcommand{\be}{\begin{equation}}
\newcommand{\ee}{\end{equation}}
\newcommand{\bea}{\begin{eqnarray}}
\newcommand{\eea}{\end{eqnarray}}
\newcommand{\bi}{\begin{itemize}}
\newcommand{\ei}{\end{itemize}}

\newcommand {\bD}{\overline{D}}

\usepackage{amscd,amsmath,amssymb}

\begin{document}
\thispagestyle{empty}
\vspace{2cm}
\begin{flushright}
hep-th/0511054 \\
[3cm]
\end{flushright}

\begin{center}
{~}\\
\vspace{3cm}
{\Large\bf $N=8$ Nonlinear Supersymmetric Mechanics}\\
\vspace{2cm}
{\large \bf S.~Bellucci${}^{a}$, A.~Beylin${}^{b}$,  S.~Krivonos${}^{b}$ and A.~Shcherbakov${}^{b}$  }\\
\vspace{2cm}
{\it ${}^a$INFN-Laboratori Nazionali di Frascati, Via E. Fermi 40,
00044 Frascati, Italy}\\
{\tt bellucci@lnf.infn.it} \\\vspace{0.5cm}
{\it ${}^b$ Bogoliubov  Laboratory of Theoretical Physics, JINR,
141980 Dubna, Russia}\\
{\tt beylin, krivonos, shcherb@theor.jinr.ru} \\ \vspace{2.5cm}
\end{center}

\begin{center}
{\bf Abstract}
\end{center}
We construct a new two-dimensional $N=8$ supersymmetric mechanics with nonlinear chiral supermultiplet. Being
intrinsically nonlinear this multiplet describes 2 physical bosonic and 8 fermionic degrees of freedom. We construct
the most general superfield action of the sigma-model type and propose its simplest extension by a Fayet-Iliopoulos
term. The most interesting property of the constructed system is a new type of geometry in the bosonic subsector, which
is different from the special K\"{a}hler one characterizing the case of the linear chiral $N=8$ supermultiplet. \hfil
\newpage

\section{Introduction}
Supersymmetric mechanics with $N=8$ supersymmetry plays a special role among extended supersymmetric theories in one
dimension. Firstly, like other theories with eight real supercharges, it admits an off-shell superfield formulation,
which simplifies its analysis. At the same time, the restrictions which $N=8$ extended supersymmetry imposes on the
geometry of the target space (see e.g. \cite{1}) are not too strong. Really speaking, the $N=8$ supersymmetric
mechanics is the highest-$N$ case among theories in one dimension which have a non-trivial geometry of the bosonic
sector. Another important property which selects just $N=8$ supersymmetry is the structure of supermultiplets. It is
shown in \cite{GR} that $N=8$ is, once again, the highest-$N$ case when the minimal supermultiplets contain $N$-bosons
and $N$-fermions. For $N>8$ supersymmetry the length of the supermultiplets grows dramatically, what makes the analysis
cumbersome.

The first detailed investigation of the one-dimensional sigma-models with $N=8$ supersymmetry is carried out in
\cite{GPS}. The relevant actions for 8 bosons and 8 fermions describe $N=8, d=1$ sigma-models with strong torsionful
octonionic-K\"{a}hler (OKT) bosonic target geometries. In \cite{DE}, a superconformal $N=8, d=1$ action is constructed
for the supermultiplet with 5 bosons and 8 fermions (see also \cite{{BMZ},{AS},{IS}}). Other versions of $N=8$
superconformal mechanics are considered in \cite{BIKL1}.

Within the off-shell superfield approach, the main ingredient one should  know is the complete list of the
corresponding off-shell supermultiplets. In \cite{ABC} the full set of linear off-shell $N=8$ supermultiplets was deduced
by joining of two off-shell $N=4$ supermultiplets and imposing the four extra supersymmetries mixing these
two $N=4$ superfields.

But the list of $N=8$ off-shell supermultiplets includes also a variety of non-linear ones. Some of these supermultiplets
may be constructed in the same way as in the case of $N=4$ supersymmetry \cite{ikl1}. However, the case of $N=8$ supersymmetry
is more complicated, as compared to the $N=4$ one, due to the existence of four non-equivalent $N=8$ superconformal
groups with numerous coset manifolds. Moreover, for some supermultiplets the constraints which follow from the method
used in \cite{ikl1} are not enough to describe irreducible supermultiplets. Indeed, even for linear $N=8$ supermultiplets \cite{ABC}
one should involve the second order (in spinor covariant derivatives) constraints which have no clear geometric meaning.
So, up to now,  we have no exhaustive list of off-shell irreducible $N=8, d=1$ supermultiplets.

Nevertheless, after completing the detailed investigation of the superfield
structure of all linear $N=8, d=1$ supermultiplets in \cite{ABC}, new variants of $N=8$ supersymmetric mechanics have
been developed \cite{{bkn1},{bks1},{sut},{bis}}.
A variety of the constructed $N=8, d=1$ systems still possesses the same property -- the metrics of the bosonic
sigma-models are conformally flat and obey $n$-dimensional Laplace equations for the systems with $n$ physical bosons.
In the case with two physical bosons this condition immediately yields a special K\"{a}hler geometry, while for 8
bosons it is equivalent to OKT geometry. The natural question then arises, whether it is possible to construct some
variant of $N=8, d=1$ supersymmetric mechanics having a different type of geometry? The answer is, for sure, positive
for the theories with 4 bosons and 8 fermions which can be obtained via dimensional reduction from any $N=2, d=4$
hyper-K\"{a}hler sigma-models. But all such theories admit off-shell formulations only in harmonic \cite{book} or
projective superspace \cite{pss}, while in ordinary superspaces the constraints describing these supermutiplets are
on-shell. Moreover, all these theories will contain  $4k$ bosons. But what about the simplest $N=8,d=1$ supermultiplet
with 2 bosons and 8 fermions? Just this case seems to be rather interesting, owing to a deep relation to $N=2, d=4$
supersymmetric Yang-Mills theory. In the present paper we propose a new nonlinear $N=8$ supermultiplet which describes
2 physical bosons and 8 fermions\footnote{We use the notation ({\bf n,N,N-n}) to describe a supermultiplet with $n$
physical bosons, $N$ fermions and $N-n$ auxiliary bosons. In this notation our supermultiplet is the ({\bf 2,8,6})
one.}. In its construction we combined the constraints describing $N=4$ nonlinear supermultiplet \cite{{ikl1},{bbkno}}
with the constraints for $N=8$ linear $({\bf 2,8,6})$ supermultiplet \cite{ABC,bkn1,bks1}. The action for this
nonlinear chiral supermultiplet (NCM) may be written in terms of $N=8$ superfields in the chiral sub-superspace  of
$N=8, d=1$ superspace. We analyze the properties of the components action and perform the Hamiltonian analysis of the
constructed system. As we expected, the bosonic metrics of the model is not a special-K\"{a}hler one.

\setcounter{equation}0
\section{N=8 Nonlinear chiral supermultiplet in superspace}
In the construction of $N=8$ NCM we will combine two ideas. One is coming from the description of $N=4$ NCM
\cite{ikl1,bbkno}, whereas the second one -- from the structure of the $({\bf 2,8,6})$ supermultiplet in $d=1$
\cite{ABC,bkn1,bks1}. Thus, in order to get the $N=8$ superfield formulation of $N=8, d=1$ NCM, we introduce a complex
scalar bosonic superfield $\cZ$ obeying the constraints
\bea
&& D^a\cZ = -\alpha\cZ \bD{}^a \cZ \; , \quad \nabla^\alpha \cZ = -\alpha\cZ \bn{}^\alpha \cZ , \label{con1} \\
&& \bn{}^\alpha \bD{}^a \cZ + \nabla^\alpha D^a \cbZ = 0. \label{con2}
\eea
Here, the covariant spinor derivatives $D^a, \bD_a, \nabla^\alpha, \bn_\beta $ are defined in the $N=8, d=1$ superspace
$\mathbb{R}^{(1|8)}$
\be
\mathbb{R}^{(1|8)} = (t, \,\theta_a,\, \bar\theta^b\; ,\vartheta_\alpha ,\, \bar\vartheta{}^\beta ) ,\quad \overline{
\left( \theta_a\right)} =\bar\theta{}^a,\; \overline{\left( \vartheta_\alpha \right)} =\bar\vartheta{}^\alpha, \quad a,
b, \alpha, \beta = 1, 2
\ee
by
\bea\label{sderiv}
&& D^{a}=\frac{\partial}{\partial\theta_{a}}+i\bar\theta{}^{a}\partial_t\,,\;
\bD_{a}=\frac{\partial}{\partial\bar\theta{}^{a}}+i\theta_{a}\partial_t\,,\quad \{D^a, \bD_b\} = 2i \delta^a_b
\partial_t , \nn &&
\nabla^{\alpha}=\frac{\partial}{\partial\vartheta_{\alpha}}+i\bar\vartheta{}^{\alpha}\partial_t\,,\;
\bn_{\alpha}=\frac{\partial}{\partial\bar\vartheta{}^{\alpha}}+i\vartheta_{\alpha}\partial_t\,,\quad \{ \nabla^\alpha,
\bn_\beta \} = 2i \delta^\alpha_\beta \partial_t .
\eea
If the real parameter $\alpha$ does not vanish, it is always possible to pass to $\alpha=1$ by a redefinition of the
superfields $\cZ, \cbZ$. So, we are left with only two essential values $\alpha=1$ and $\alpha=0$. In what follows we
will put $\alpha=1$.

With $\alpha=0$ the constraints \p{con1},\p{con2} describe the linear $N=8$ $({\bf 2,8,6})$ supermultiplet \cite{ABC}.
On the other hand, the subset \p{con1} constitutes two copies of the $N=4$ NCM constraints restricting the $\theta$-
and $\vartheta$-dependence of the $N=8$ superfields $\cZ, \cbZ$.

The component structure of the $N=8$ superfield $\cZ$, implied by \p{con1}, \p{con2}, is a bit involved in comparison
with the $N=4$ NCM or $\alpha=0$ cases. In order to define it, let us firstly consider the constraints \p{con1}. It
immediately follows from \p{con1} that the derivatives $D^a$ and $\nabla^\alpha$ of the superfield $\cZ$ (or $\bD{}^a,
\bn{}^\alpha$ of $\cbZ$) can be expressed as $\bD{}^a, \bn{}^\alpha$ (or $D^a, \nabla^\alpha$) derivatives of the same
superfield. Therefore, as in the case of chiral superfields, only the components appearing in the $\bar\theta{}^a,
\bar\vartheta{}^\alpha$ expansion of $\cZ$ and the $\theta{}^a, \vartheta{}^\alpha$ expansion of the $\cbZ$ superfield
are independent. Let us define these components as follows\footnote{All implicit summations go from ``up-left'' to
``down-right'', e.g., $\psi\bpsi \equiv \psi^i\bpsi_i$, $\psi^2 \equiv \psi^i\psi_i$,  etc.}:
\be\label{comp}
\ba{ll}
\mbox{Bosonic components}&\\[2mm]
 z=\cZ, & \bz=\cbZ ,\\
 A=-i\bD{}^2\cZ ,& \bar{A}=-iD^2\cbZ ,\\
 B=-i\bn{}^2\cZ ,& \bar{B}=-i\nabla^2\cbZ,\\
 Y^{a\alpha}=\bD{}^{a}\bn{}^{\alpha}\cZ ,& \bar{Y}{}^{a\alpha}=-D^{a}\nabla^{\alpha}\cbZ ,\\
 X=\bD{}^2\bn{}^2\cZ ,&\bar X = D^2 \nabla^2\cbZ
\ea\qquad \ba{ll}
\mbox{Fermionic components}&\\[2mm]
 \psi_a= \bD_a\cZ ,& \bpsi_a=-D_a\cbZ,\\
 \xi_\alpha=\bn_\alpha \cZ ,& \bxi_\alpha=-\nabla_\alpha\cbZ ,\\
\tau_\alpha = \bD{}^2 \bn_\alpha \cZ ,& \bar\tau_\alpha= D^2\nabla_\alpha\cbZ ,\\
 \sigma_a=\bn{}^2 \bD_a\cZ ,& \bar\sigma_a=\nabla^2 D_a\cbZ ,\\
 \\
\ea
\ee
where the right hand side of each expression is supposed to be taken with $\theta=\vartheta=0$. Thus, the first part of
our constraints \p{con1} leaves in the $N=8$ superfields $\cZ, \cbZ$ 16 bosonic and 16 fermionic components.

Now it is time to consider the constraints \p{con2}. Besides the evident reality condition on the $Y^{a\alpha}$
\be\label{Y}
Y^{a\alpha}=\bar{Y}{}^{a\alpha},
\ee
they put the following restriction on the components \p{comp}:
\be\label{bos}
\ba{l} X=16\ddot\bz - 4 \partial_t(\bz\bar{A} + \bz\bar{B} -i \bpsi{}^2-i\bxi{}^2) +\bz \left( 4 Y^{a\alpha}Y_{a\alpha}
+ 2\bar{A}\bar{B}  -
 \bz \bar X\right) \\[1mm]
\qquad -2i \left[ \bar{B}\bpsi{}^2 +\bar{A} \bxi{}^2+4i\bxi{}^\alpha\bpsi^a Y_{a\alpha}+
     2i\bz(\bpsi{}^a\bar\sigma_a+\bxi{}^\alpha\bar\tau_\alpha )\right], \\[2mm]
\bar  X=16\ddot z - 4 \partial_t( zA + zB -i \psi{}^2-i\xi{}^2) + z \left( 4 Y^{a\alpha}Y_{a\alpha} + 2 AB  -
 zX  \right) \\[1mm]
\qquad -2i \left[ B\psi{}^2 + A \xi{}^2-4i\xi{}^\alpha\psi^a Y_{a\alpha}-
     2i z (\psi{}^a \sigma_a+\xi{}^\alpha\tau_\alpha )\right], \\[2mm]
 4\dot{\bar{A}}- \left[ \bar{A}\bar{B}+2(\bpsi{}^a\bar{\sigma}_a+\bxi^\alpha\bar\tau_\alpha) -\bz\bar X\right]=
       4\dot B - \left[ A B- 2(\psi^a\sigma_a+ \xi^\alpha\tau_\alpha)-z X\right],\\[2mm]
 4\dot{A}- \left[ AB-2(\psi{}^a\sigma_a+\xi^\alpha\tau_\alpha) -z X\right]=
       4\dot {\bar B} - \left[ \bar{A}\bar{B}+ 2(\bpsi{}^a\bar\sigma_a+ \bxi{}^\alpha\bar\tau_\alpha)-\bz \bar{X}\right],
\ea
\ee
\be\label{ferm}
\ba{l}
 \tau^\alpha= -4i\dot\bxi{}^\alpha +\left( 2Y^{a\alpha}\bpsi_a-\bar\tau{}^\alpha \bz +i\bar{A}\bxi{}^\alpha\right),\;
\bar\tau{}^\alpha= -4i\dot\xi{}^\alpha -\left( 2Y^{a\alpha}\psi_a-\tau{}^\alpha z -i A\xi{}^\alpha\right), \\[1mm]
  \sigma^a= -4i\dot\bpsi{}^a -\left( 2Y^{a\alpha}\bxi_\alpha+\bar\sigma{}^a\bz-i\bar{B}\bpsi{}^a\right),\;
\bar\sigma{}^a= -4i\dot\psi{}^a +\left( 2Y^{a\alpha}\xi_\alpha+\sigma{}^a z+i B \psi{}^a\right).
 \ea
\ee

The first two equations in \p{bos} and eq.\p{ferm} express the auxiliary bosonic components $X, \bar X$ and fermionic
ones $\tau^\alpha, \bar\tau{}^\alpha, \sigma^\alpha, \bar\sigma{}^\alpha$ in terms of physical bosons and fermions and
auxiliary bosons $A, \bar A, B, \bar B$. For these auxiliary bosons we have differential equations (two last lines in
\p{bos}) which should be somehow solved. In the $\alpha=0$ case, which corresponds to discarding all nonlinear terms,
these equations read
\be\label{lin}
\partial_t \left( {\bar A} -B\right) =0, \quad \partial_t \left( A -{\bar B}\right) =0
\ee
and may be easily solved. In $\alpha \neq 0$ case we can also find a proper solution. In order to do this, we use the
following ansatz:
\be\label{anz}
A=\bB+f,\qquad \bA=B-f, \qquad \bar f=-f
\ee
inspired by $\alpha=0$ limit of the constraints. Substituting \p{anz} into the eqs.(\ref{bos}) we get the following
equation:
\be
(B+\bB)\left[ {z}\dot\bz - \bz\dot{z} -\frac{i}{2} (\xi\bxi+\psi\bpsi) + \frac{1}{4}(1+{z}\bz)f \right] -4\partial_t
\left[ {z}\dot\bz - \bz\dot{z} -\frac{i}{2} (\xi\bxi+\psi\bpsi)+\frac{1}{4}(1+{z}\bz)f \right]=0
\ee
with the obvious solution
\be\label{f}
f=4 \frac{\bz\dot{z}-{z}\dot\bz}{1+{z}\bz}+2i\frac{\xi\bxi+\psi\bpsi}{1+{z}\bz}.
\ee
Thus, we have only six auxiliary fields, $B,\bar{B}, Y^{a\alpha}$ and therefore our nonlinear supermultiplet is just
the $(2,8,6)$ one.

It is worth noticing that differential constraints of the considered type may be exactly solved (as in \p{f}) only in
one dimension. In all other cases we have to insert these differential constraints \p{bos} with Lagrange multipliers in
the proper action. Due to the differential nature of the constraints, these Lagrangian multipliers will be vectors with
respect to the corresponding Lorentz group. Varying over the auxiliary fields we will express them through
field-strengths constructed from the Lagrangian multipliers. Thus, the corresponding theory will contain gauge fields.
In one dimension we may also plug these constraints in the proper action. In contrast with higher dimensional cases
this gives rise to a $(4,8,4)$ supermultiplet in full analogy with the $\alpha=0$ case \cite{bks1}. We will not
consider this option here.

\setcounter{equation}0
\section{Action and Hamiltonian}
Now one can write the most general $N=8$ sigma-model type  action in $N=8$ superspace\footnote{We use
the convention $\int d^2 \theta = \frac {1}{4} D^{i} D_i$, $\int d^2 \bar\theta = \frac {1}{4} \bD_i \bD^i$.}
\be\label{action1}
S =  \int\! dt d^2 \bar\theta d^2 \bar\vartheta\; \cF(\cZ) + \int\! dt
    d^2\theta d^2 \vartheta\; \cbF (\cbZ) \;.
\ee
Here  $\cF (\cZ)$ and $\cbF(\cbZ)$ are arbitrary holomorphic functions depending only on $\cZ$ and $\cbZ$,
respectively. Let us stress that in \p{action1} we integrate over (anti)chiral superspace while our superfields are not
(anti)chiral. Usually such a trick is forbidden, because the action fails to be invariant under supersymmetry. But for
the NCM with the constraints \p{con1}, \p{con2} the action \p{action1} is  invariant with respect to the full $N=8$
supersymmetry. Indeed, the supersymmetry transformations of the integrand of, for example, the first term in
\p{action1} which seems to break supersymmetry read
\be\label{dok1}
\delta \cF(\cZ) \sim -\epsilon^a D_a \cF(\cZ) = \epsilon^a \cF_{\cZ} \cZ \bD_a \cZ.
\ee
It is evident that the right hand side can always be represented as a $\bD$-derivative of a function of $\cZ$. Hence,
the variation in \p{dok1} disappears after integration over $d^2 \bar\theta$ and therefore the action \p{action1} is
invariant with respect to the full $N=8$ supersymmetry.

After integrating in \p{action1} over the Grassmann variables and excluding the auxiliary fields $B, \bar{B},
Y^{a\alpha}$ by their equations of motion, we will get the action in terms of physical components
\bea\label{fullact}
S&=& \int \! d t \Biggl[ g \dz \dbz +
\frac{i}{4} g(1 + z \bz) \left ( \dot{\bxi} \xi - \dot{\xi} \bxi +
\dot{\bpsi} \psi - \dot{\psi} \bpsi \right) +\nn &+&\frac{i}{4} g \left( \dbz
(\xi^2 + \psi^2) + \dz(\bxi^2 + \bpsi^2) \right) + \frac i4 (\xi\bxi+\psi\bpsi)\left(\dbz
\bar\diff_z\tg - \dz \diff_z\tg \right) - \nn &-&\frac{1}{16}\left ( \diff_z \diff_z \g + \frac{2
\bz}{1 + z \bz} \diff_z \g \right) \psi^2 \xi^2
- \frac{1}{16}\left ( \bar\diff_z \bar\diff_z \g + \frac{2 z}{1 + z \bz} \bar\diff_z \g \right) \bpsi^2 \bxi^2 + \\
&+& \frac{g}{8} \left( \xi^2\bxi^2 + \psi^2\bpsi^2 + \xi^2 \bpsi^2 + \bxi^2 \psi^2 \right)
+ \frac{3}{16}\frac{(\diff_z \g)^2}{\g} \psi^2 \xi^2 + \frac{3}{16}\frac{(\bar\diff_z \g)^2}{\g} \bpsi^2 \bxi^2 + \nn
&+& \frac{1}{16}\frac{\diff_z \g \bar\diff_z \g}{\g} \left( \psi^2\bpsi^2 + \xi^2\bxi^2 - 4 \xi\bxi\psi\bpsi \right)
+ \frac14\frac{\bar\diff_z \g}{1 + z \bz}\left( \xi\bxi\psi^2 + \bxi^2\psi\bpsi \right)
- \frac14\frac{\diff_z \g}{1 + z \bz} \left( \psi\bpsi\xi^2 + \psi^2\xi\bxi \right) \nonumber
\Biggr ].
\eea
Here, we define the metric $g$
\be\label{defs}
g({z},\bz)=  \frac{F''+\bF''}{(1+{z}\bz)^2}-
       2\frac{F'\bz+\bF'{z}}{(1+{z}\bz)^3} =\frac{\partial^2 K}{\partial z \partial \bz}, \qquad
       K=\frac{F'\bz+\bF'{z}}{(1+{z}\bz)},
\ee
and the auxiliary functions $\tilde g$ and $\gamma$
\be\label{defs1}
\tilde g = (1+ z\bz)g,\quad \gamma= (1+ z\bz)^2 g \;.
\ee
Thus, we see that the bosonic subsector of the action \p{fullact} describes a K\"{a}hler sigma-model with the metric
\p{defs}. In contrast to $\alpha=0$ case with the special K\"{a}hler metric \cite{ikl1}, the NCM supermultiplet
describes the sigma-model with a different metric. It is not immediately clear which type of geometry corresponds to
the metric \p{defs}. All we can say now is that the metric $g$ is the solution to the following equation:
\be\label{gamma}
\frac{\partial^2}{\partial z \partial \bz}\left[ (1+z\bz)^2 g\right] = -2 g.
\ee
Let us also note that, using the gauge freedom, the K\"{a}hler potential \p{defs} can be rewritten in the
following equivalent form:
\be
K=\frac{F'\bz+\bF'{z}}{(1+{z}\bz)}-\frac{F'}{z} -\frac{\bF'}{\bz} = -\frac{H +{\overline H}{}'}{(1+z\bz)},
\ee
where
\be
H(z)\equiv \frac{F'}{z}, \quad {\overline H}(\bz) \equiv \frac{\bF'}{\bz}.
\ee
It is interesting that the simplest case with
\be
F(z) = z^2
\ee
which corresponds to the free action for the linear chiral supermultiplet  gives the following metric $g$ for the NCM:
\be
g = 4 \;\left(\frac{1-z\bz}{1+z\bz}\right)\; \frac{1}{(1+z\bz)^2}
\ee
which looks as a metric on the sphere deformed by the factor $\left(\frac{1-z\bz}{1+z\bz}\right)$.

One may easily check that in the limit $\psi=0$ (or $\xi=0$) the action \p{fullact} goes into the action of the $N=4$ NCM
\cite{bbkno}.

Now it becomes clear that the net effect of using the $N=8$ NCM with respect
to standard ones is a new type of metric in the bosonic subsector which is not anymore of the special-K\"{a}hler kind.

In the next Section we will clarify the new features of the sigma-model with the NCM in the Hamiltonian formalism.

\setcounter{equation}0
\section{N=8 NCM: Hamiltonian approach}
In order to find the classical Hamiltonian, we start from the action \p{fullact} and first of all define the momenta
$p, \bp, \pi^{(\psi)}_a,\bar\pi^{(\psi)a} \pi^{(\xi)}_\alpha ,\bar\pi^{(\xi)\alpha}$ as
\bea\label{momenta}
&&p=g\dot\bz+\frac{i}{4}\left(g(\bpsi^2+\bxi^2)-\diff_z\tg (\psi\bpsi+\xi\bxi)\right),\\
&&\bp=g\dot z+\frac{i}{4}\left(g(\psi^2+\xi^2)+\bar\diff_z\tg (\psi\bpsi+\xi\bxi)\right),\\
&&\pi^{(\psi)}_a=-\frac{i}{4} \tg\bpsi_a,\qquad
\bar\pi^{(\psi)a}=-\frac{i}{4} \tg \psi^a, \\
&&\pi^{(\xi)}_\alpha=-\frac{i}{4}\tg\bxi_\alpha,\qquad
\bar\pi^{(\xi)\alpha}=-\frac{i}{4}\tg \xi^\alpha
\eea
with the metric $g$ and auxiliary expression $\tg$ given by eqs. \p{defs} and \p{defs1}. Because of the presence of the
second-class constraints
\bea\label{2cconstr}
&&  \chi^{(\psi)}_a=\pi^{(\psi)}_a+\frac i4\tg\bpsi_a,\;
 \bar\chi{}^{(\psi)a}=\bar\pi{}^{(\psi)a}+\frac i4\tg\psi^a,\;
 \chi^{(\xi)}_\alpha=\pi^{(\xi)}_\alpha+\frac i4\tg\bxi_\alpha,\;
 \bar\chi{}^{(\xi)\alpha}=\bar\pi{}^{(\xi)\alpha}+\frac i4\tg\xi^\alpha, \nn
&&
\left\{ \chi^{(\psi)}_a, \bar\chi{}^{(\psi)b}\right\}=-\frac{i}{2}\tg\delta_a^b ,\quad
\left\{ \chi^{(\xi)}_\alpha, \bar\chi{}^{(\xi)\beta}\right\}=-\frac{i}{2}\tg\delta_\alpha^\beta ,
\eea
we will pass to Dirac brackets
\bea\label{Dbr}
&&\{ z,\hp \}=1,\quad
     \{\bz,\bhp \}=1, \nn
&&      \{ \hp,\,\bhp \}=-\frac i2 \frac N{g(1+z\bz)}(\psi\bpsi+\xi\bxi)
    -\frac i4 \frac M{1+z\bz}(\psi^2+\xi^2)
    +\frac i4 \frac{\bar M}{1+z\bz}(\bpsi^2+\bxi^2),\nn
&& \{\hp,\psi^a \} = \frac{\diff_z \tg}{\tg}\psi^a-
    \frac{\bpsi^a}{1+z\bz}, \qquad
 \{\hp,\xi^\alpha \} = \frac{\diff_z \tg}{\tg}\xi^\alpha-
    \frac{\bxi^\alpha}{1+z\bz},\nn
&& \{\bhp,\bpsi_a \} = \frac{\bar\diff_z \tg}{\tg}\bpsi_a+
    \frac{\psi_a}{1+z\bz},\qquad
 \{\bhp,\bxi_\alpha \} = \frac{\bar\diff_z \tg}{\tg}\bxi_\alpha+
    \frac{\xi_\alpha}{1+z\bz},\nn
&& \{\psi^a,\bpsi_b\}=-\frac {2i}{\tg} \delta^a_b,\qquad
 \{\xi^\alpha,\bxi_\beta\}=-\frac{2i}{\tg}\delta^\alpha_\beta.
\eea
Here we have introduced new bosonic momenta $\hp, \bhp$ as
\be\label{bmom}
\hp \equiv g{\dot{\bz}}=p-\frac i4 g (\bpsi^2+\bxi^2)+
    \frac i4 \diff_z \tg(\psi\bpsi +\xi\bxi),\quad
\bhp \equiv g\dot{z}= \bp-\frac i4 g (\psi^2+\xi^2)-
    \frac i4 \bar\diff_z \tg(\psi\bpsi +\xi\bxi),
\ee
and defined $M, \bar M$ and $N$
\bea\label{MN}
&&M \equiv (1+z\bz)\diff_z g +2 g \bz, \qquad
\bar M \equiv (1+z\bz)\bar\diff_z g +2 g z ,\nn
&&N=2g^2(1+2z\bz)+2g(z\diff_z g+\bz\bar\diff_z g)(1+z\bz)
    +\diff_z g\bar\diff_z g (1+z\bz)^2 .
\eea
Being expressed in terms of the new momenta \p{bmom}, the Hamiltonian takes a canonical form
\bea\label{Ham}
H&=&\frac 1g \hp\bhp-\frac 18 g (\psi^2\bxi^2+\bpsi^2\xi^2)
    +\frac 14 \psi\bpsi(M\xi^2-\bar M\bxi^2)
    +\frac 14 \xi\bxi(M\psi^2-\bar M\bpsi^2)\nn
&+&\frac1{16g}\left(4M\bar M \psi\bpsi\xi\bxi
    + K\psi^2\xi^2+\bar K\bpsi^2\bxi^2
    - (\xi^2\bxi^2+\psi^2\bpsi^2)N\right)
\eea
with
\be\label{K}
K=-6g\bz\diff_z g (1+z\bz) + (g\diff_z \diff_z g -3(\diff_z g)^2)(1+z\bz)^2 - 6 g^2\bz^2 .
\ee
Now one can check that the supercharges $Q_a, {\bar Q}_a, S_\alpha, {\bar S}_\alpha$
\bea\label{chrg}
Q_a&=&\hp\psi_a -\bhp\bz\bpsi_a+\frac i4 g\psi_a (\bpsi^2+2\bz\xi\bxi)
    -\frac i4 g\bpsi_a (2\xi\bxi+\bz\psi^2)\nn
    &-&\frac i4\bpsi_a\bxi^2\left((1+z\bz)\bar\diff_z g +2 g z\right)
    -\frac i4\psi_a\xi^2\bz \left((1+z\bz)\diff_z g +2 g\bz\right)\nn
S_\alpha&=&\hp\xi_\alpha -\bhp\bz\bxi_\alpha+\frac i4 g\xi_\alpha (\bxi^2+2\bz\psi\bpsi)
    -\frac i4 g\bxi_\alpha (2\psi\bpsi+\bz\xi^2)\nn
    &-&\frac i4\bxi_\alpha\bpsi^2\left((1+z\bz)\bar\diff_z g +2 g z\right)
    -\frac i4\xi_\alpha\psi^2\bz \left((1+z\bz)\diff_z g +2 g\bz\right)\\
{\bar S}{}^\alpha&=& \left( S_\alpha \right)^\dagger, \quad {\bar Q}^a=\left( Q_a \right)^\dagger,
\eea
form, together with the Hamiltonian $H$ \p{Ham}, the $N=8$ Poincar\`{e} superalgebra
\be\label{Alg}
\{ Q_a,\bar Q_b\}=-2i\epsilon_{ab}H,\qquad
\{ S_\alpha,\bar S_\beta\}=-2i\epsilon_{\alpha\beta}H.
\ee
Thus, we conclude that $N=8$ supersymmetric mechanics with NCM corresponds to the rather non-trivial construction of
supercharges and describes the sigma-model with the metric $g$ in its bosonic sector.

\setcounter{equation}0
\section{N=8 NCM: Potential terms}
Up to now we considered only the action of the sigma-model type. In full analogy with $\alpha=0$ case of \cite{bks1}
one may try to add Fayet-Iliopoulos terms to our action \p{action1} in order to generate some potential terms. In the
NCM supermultiplet we have two scalar auxiliary fields $B, \bar B$, but due to the nonlinear nature of the NCM only
their sum transforms through a full time derivative. Thus, we consider the following action:
\be\label{fi}
\tilde{S}=S + m\int \! dt \left( B+\bar B \right) \;.
\ee
In physical components the action \p{fi} reads
\be\label{fi1}
\tilde{S} =S+ \int \! dt \left[ \frac{im}{4\g} \left(\diff_z\g(\psi^2+\xi^2)+\bar\diff_z \g(\bpsi^2+\bxi^2)\right)
    - \frac{m^2}{\g} \right] \;,
\ee
with $\g$ defined in \p{defs1}.

The supercharges \p{chrg} and Hamiltonian are modified as follows:
\bea\label{fi2}
&& {\tilde Q}_a=Q_a+\frac{m}{1+z\bz}(\bpsi_a + \psi_a \bz),\quad {\tilde S}_\alpha=S_\alpha
+ \frac{m}{1+z\bz}(\bxi_\alpha + \xi_\alpha \bz), \nn
&& {\tilde H} =H - \cfrac{imM}{g(1+z\bz)}(\psi^2+\xi^2) - \cfrac{im\bar M}{g(1+z\bz)}(\bpsi^2+\bxi^2)
    + \cfrac{m^2}{g(1+z\bz)^2} \;,
\eea
while the Dirac brackets \p{Dbr} remain unchanged. It is interesting to stress that even for the trivial metric $g=1$
the Hamiltonian still contains the potential term.

\section{Conclusion}
In the present paper we constructed $N=8$ supersymmetric mechanics with the nonlinear chiral supermultiplet. The main
interesting peculiarity of the constructed system is the appearance of the ``modified'' metric, which is not of the
special K\"{a}hler type. We are still unable to identify this new geometry which is selected as the solution of the
equation
$$
\frac{\partial^2}{\partial z \partial \bz}\left[ (1+z\bz)^2 g\right] = -2 g.
$$
We also introduced the Fayet-Iliopoulos term, in order to get potential terms in the action. Unfortunately, no
interaction with the magnetic field could be implemented.\footnote{Issues related to the possibility to exactly solve
extended supersymmetric mechanics systems after inclusion of a constant magnetic field were treated for $N=4$ in
earlier papers \cite{BN,BNY}.}

These results should be regarded as preparatory for a more detailed study of supersymmetric mechanics
with nonlinear supermultiplets. It is clear now that the unique way to have an interesting geometry in
the bosonic target space is to use nonlinear supermultiplets. In this respect it seems interesting
to convert some of the auxiliary bosons into physical ones. In this way one could expect to find
some new nonlinear supermultiplets with more than 2 physical bosons and, therefore, one could be able to construct
a supersymmetric systems with a new type geometry of the target space.

Another related question concerns the possibility to extend NCM to higher dimensions. Indeed, it seems that it should be
possible to extend the defining relations to higher dimensions, at least those where the spinor
covariant derivative $D^i_\alpha$ and its conjugated expression ${\bar D}^i_\alpha$
have the same Lorentz indices. The first  case is evidently  $N=4, d=3$ supersymmetry. Certain steps towards the
construction of the nonlinear $N=4, d=3$ vector supermultiplet will be presented in a forthcoming paper \cite{new}.

\section*{Acknowledgements}
We wish to thank A.P.~Isaev, E.A.~Ivanov and A.P.~Nersessian for valuable discussions. We are indebted to the Referee for
useful remarks.

This research was partially supported by the European Community's Marie Curie Research Training Network under contract
MRTN-CT-2004-005104 Forces Universe, and by the INTAS-00-00254 grant. S.K. and A.S. thank INFN --- Laboratori Nazionali
di Frascati  for the warm hospitality extended to them during the course of this work.

\end{document}